\documentclass[9pt,twocolumn,twoside]{opticajnl}
\journal{opticajournal} % use for journal or Optica Open submissions

\setboolean{shortarticle}{true}
\usepackage{lineno}
\usepackage[none]{hyphenat}
\usepackage{braket}
\usepackage[font=small,labelfont=bf,justification=justified]{caption}
\usepackage{hyperref}

\title{Dispersion effects in thermal emission from temporal metamaterials: High-frequency cut-offs}

\author[$\,\!$]{Amaia Vertiz-Conde}
\author[*]{I\~nigo Liberal}
\author[$\dagger$]{J. Enrique V\'azquez-Lozano}

\affil[$\,\!$]{Department of Electrical, Electronic and Communications Engineering, Institute of Smart Cities, Universidad Pública de Navarra, 31006 Pamplona, Spain}

\affil[*]{inigo.liberal@unavarra.es}
\affil[$\dagger$]{enrique.vazquez@unavarra.es}

\begin{abstract}
The latest breakthroughs in time-varying photonics are fueling novel thermal emission phenomena, for example, showing that the dynamic amplification of quantum vacuum fluctuations, induced by the time-modulation of material properties, enables a mechanism to surpass the black-body spectrum. So far, this issue has only been investigated under the assumption of non-dispersive time-modulations. In this work, we identify the existence of a non-physical diverging behavior in the time-modulated emission spectra at high frequencies, and prove that it is actually attributed to the simplistic assumption of a non-dispersive (temporally local) response of the time-modulation associated with memory-less systems. Accordingly, we upgrade the theoretical formalism by introducing a dispersive response function, showing that it leads to a high-frequency cut-off, thereby eliminating the divergence and hence allowing for the proper computation of the emission spectra of time-modulated materials.
\end{abstract}

\setboolean{displaycopyright}{true} % Do not include copyright or licensing information in submission.

\hypersetup
{
	pdfauthor={
		Amaia Vertiz-Conde (vertiz.138863@e.unavarra.es),
		I\~nigo Liberal (inigo.liberal@unavarra.es) \&
		J. Enrique V\'azquez-Lozano (enrique.vazquez@unavarra.es)},
	pdfsubject={
		Department of Electrical, Electronic and Communications Engineering, Institute of Smart Cities (ISC), Universidad P\'ublica de Navarra (UPNA)},
	pdftitle={
		Dispersion effects in thermal emission from temporal metamaterials: High-frequency cut-offs},
	pdfkeywords={thermal emission, temporal metamaterials, nanophotonics, quantum optics, black-body radiation, time-modulated dispersion}
}

\begin{document}

\maketitle
\sloppy

In recent decades, significant advancements in nanophotonics and materials science have sparked renewed interest in the study of thermal radiation~\cite{Cuevas2018,Li2018,Baranov2019,Li2021}, laying down the basis for what is now recognized as thermal emission engineering~\cite{VazquezLozano2024}. Similar to nanophotonic engineering~\cite{Boriskina2017}, the primary objective of thermal emission engineering is to control and manipulate the coherence properties of thermal radiation, thereby enabling mechanisms for tuning and enhancing thermal emission features such as the spectral bandwidth~\cite{Liu2011}, the directivity~\cite{Greffet2002}, and the degree of polarization~\cite{Wang2023}. To this aim, most of the practical implementations so far carried out have been based on passive approaches relying upon the use of photonic nanostructures. In this vein, the latest qualitative leap has come with the introduction of active approaches, providing with dynamic control of thermal emission features~\cite{Picardi2023}.

Yet, only very recently, and inspired by the burgeoning concept of temporal metamaterials (often simply referred to as time-varying or time-modulated media)~\cite{Engheta2021,Galiffi2022,Yuan2022,Engheta2023}, it has been put forward a comprehensive theoretical framework to address thermal emission from temporal metamaterials~\cite{Yu2023,Yu2024,VazquezLozan02023A}. Notably, according to the formulation based on macroscopic quantum electrodynamics (QED)~\cite{VazquezLozan02023A}, the temporal modulation of material properties results in extraordinary far- and near-field thermal emission features~\cite{VazquezLozan02023B}, including the potential to exceed the black-body emission spectrum set by both Planck's and Kirchhoff's radiation laws. This remarkable effect, particularly occurring at the epsilon-near-zero (ENZ) frequency range of the material, is attributed to the dynamical amplification of zero-point quantum vacuum fluctuations~\cite{Liberal2017,Liberal2018}, so that it might not be properly described by classical theory~\cite{Liberal2024}.

Likely for the sake of simplicity, the vast majority of previous works in this emerging field have routinely considered temporal modulations with an instantaneous response~\cite{VazquezLozan02023A,Ganfornina2023,Galiffi2022,Yuan2022,Sloan2021,Pendry2022}, so that the linear response function is typically characterized by means of a time-varying susceptibility displaying a temporally-local (instantaneous) dispersion, i.e., $\Delta\chi({\bf r},t,\tau) = \Delta\tilde{\chi}({\bf r},t)\delta{[t-\tau]}$. Particularly concerning the quantum domain, inasmuch as it allows for the direct utilization of equal-time commutation relations~\cite{Vogel,Scheel2008}, this assumption greatly facilitates the mathematical derivation of the Heisenberg equations of motion for the polaritonic operators~\cite{Rivera2020}, and hence the current density and electromagnetic (EM) field operators as well as the ensuing correlations. However, this approach undertakes significant limitations in fundamental physical issues~\cite{Koutserimpas2024,Sloan2024}, which is fostering an increasingly growing interest in addressing dispersion effects in~time-varying~media~\cite{Horsley2023,Horsley2024}.

In this work, by looking into the thermal emission spectra of time-modulated materials, we find out that they display an unphysical divergence at the high-frequency limit, and demonstrate that such an unbounded spectral behavior is inherently linked to the assumption of non-dispersive (or~temporally-local) time-modulation. Accordingly, we extend the theoretical formalism to encompass more realistic scenarios by incorporating the memory-time as an additional parameter. Our analysis reveals that the introduction of such a memory time into the kernel response function brings about a spectral cut-off at high frequencies. Besides eliminating the diverging behavior, thus allowing for the proper calculation of the emission spectra of temporal metamaterials, our findings also suggest a pathway for experimentally testing intrinsic material features concerning their response to external stimuli.

\begin{figure}[t!]
	\centering
	\includegraphics[width=1\linewidth]{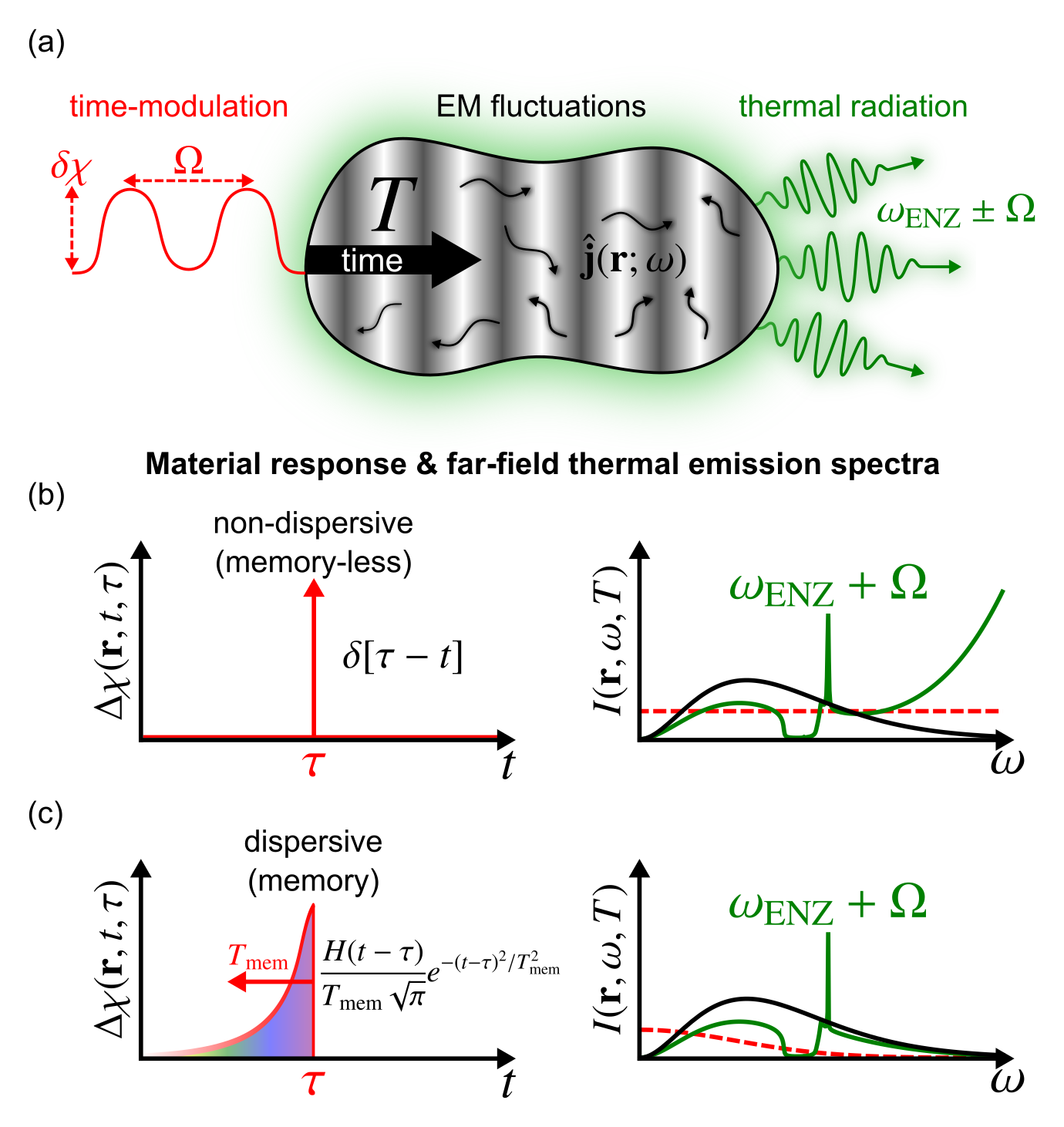}
	\caption{(a) Schematic representation of the emission of thermal radiation from EM fluctuations of a macroscopic body at temperature $T$ whose material properties are modulated under a temporal profile characterized in terms of the strength, $\delta\chi$, and the frequency, $\Omega$. (b)~Non-dispersive material response to the time-modulation yield the far-field emission spectra to diverge in the high-frequency limit. (c)~Introducing the dispersion into the material response leads to a cut-off at high-frequencies that eliminates the spectral divergence.}
	\label{Fig_01}
\end{figure}

Thermal emission has traditionally been addressed from a semiclassical approach based on fluctuational electrodynamics~\cite{Rytov}, from which the emission spectra can be simply calculated as the field density correlations, ultimately determined by the fluctuation-dissipation theorem~(FDT)~\cite{Kubo1966}:
\begin{equation}
\!\!\braket{{\bf j}^*({\bf r};\omega)\!\cdot\!{\bf j}({\bf r}';\omega')}_{\rm th}\!=\!4\pi\varepsilon_0\varepsilon''({\bf r},\omega)\hbar\omega^2\Theta(\omega,T)\delta_{{\bf r}-{\bf r}'}\delta_{\omega-\omega'},\!\!
\label{Eq01}
\end{equation}
where the brackets $\braket{\cdots}_{\rm th}$ denote a thermal ensemble average, $\Theta(\omega,T)=[e^{\hbar\omega/(k_BT)}-1]^{-1}$ stands for the black-body's photon distribution, and $\varepsilon({\bf r},\omega)=\varepsilon'({\bf r},\omega)+i\varepsilon''({\bf r},\omega)$ is the permittivity. However, in dealing with time-varying media [see~\hyperref[Fig_01]{Fig.~1(a)}], one necessarily has to take into account the occurrence of both thermal and quantum vacuum EM fluctuations, thereby requiring the introduction of a purely quantum formalism. From a first-principles approach based on macroscopic QED~\cite{Vogel,Scheel2008}, it has recently been put forward a perturbative model describing the dynamical behavior of a time-modulated quantum photonic system~\cite{VazquezLozan02023A}. It essentially consists in a Hamiltonian expressed as $\mathcal{H}=\mathcal{H}_{\rm 0}+\mathcal{H}_{\rm T}$, where $\mathcal{H}_{\rm 0}$ stands for the ground contribution characterizing the polaritons (i.e., the light-matter coupled states) hosted by the macroscopic body without time-modulation:
\begin{equation}
\hat{\mathcal{H}}_{\rm 0}=\int{d^3{\bf r}\int_{0}^{+\infty}{d\omega_f \hbar\omega_f \hat{\bf f}^\dagger({\bf r},\omega_f;t)\cdot\hat{\bf f}({\bf r},\omega_f;t)}},
\label{Eq02}
\end{equation}
and $\mathcal{H}_{\rm T}$ represents the interaction term, accounting for the perturbation yielded by the time-modulation described as a polarization field induced by an external electric field:
\begin{equation}
\hat{\mathcal{H}}_{\rm T}=-\int{d^3{\bf r}\hat{\boldsymbol{\mathcal{P}}}({\bf r};t)\cdot\hat{\boldsymbol{\mathcal{E}}}({\bf r};t)}.
\label{Eq03}
\end{equation}
The polarization field operator is generally expressed as $\hat{\boldsymbol{\mathcal{P}}}({\bf r};t)\equiv\int_{0}^{t}{d\tau \Delta\chi({\bf r},t,\tau)\hat{\boldsymbol{\mathcal{E}}}({\bf r};\tau)}$, with $\Delta\chi({\bf r},t,\tau)$ being the susceptibility, and $\hat{\boldsymbol{\mathcal{E}}}({\bf r};t)=\hat{\boldsymbol{\mathcal{E}}}^{(+)}({\bf r};t)+\hat{\boldsymbol{\mathcal{E}}}^{(-)}({\bf r};t)$ is the electric field operator, whose positive-frequency component reads as:
\begin{equation}
\hat{\boldsymbol{\mathcal{E}}}^{(+)}({\bf r};t)=\int{d^3{\bf r}'\int{d\omega_f {\bf G}_{\rm E}({\bf r},{\bf r}',\omega_f)\cdot\hat{\bf f}({\bf r}',\omega_f;t)}},
\label{Eq04}
\end{equation}
where ${\bf G}_{\rm E}({\bf r},{\bf r}',\omega_f)=i\sqrt{\hbar/\pi\varepsilon_0}[\omega_f/c]^2\sqrt{\varepsilon''({\bf r}',\omega_f)}{\bf G}({\bf r},{\bf r}',\omega_f)$ is the response function characterizing the background medium, depending in turn on the dyadic Green's function of the unmodulated material, ${\bf G}({\bf r},{\bf r}',\omega_f)$, and noticing that $\hat{\boldsymbol{\mathcal{E}}}^{(-)}({\bf r};t)=[\hat{\boldsymbol{\mathcal{E}}}^{(+)}({\bf r};t)]^\dagger$. Akin to the semiclassical approach based on fluctuational electrodynamics, from this quantum approach the thermal emission spectrum can be straightforwardly obtained from the electric field correlations:
\begin{equation}
I({\bf r},\omega,T)=\braket{[\hat{\boldsymbol{E}}^{(+)}({\bf r};\omega)]^\dagger\cdot\hat{\boldsymbol{E}}^{(+)}({\bf r};\omega) }_{\rm th},
\label{Eq05}
\end{equation}
where $\hat{\boldsymbol{E}}^{(+)}({\bf r};\omega)=\mathcal{L}_{\omega}{[\hat{\boldsymbol{\mathcal{E}}}^{(+)}({\bf r};t)]}$ is the Laplace's transform of the electric field operator given in \eqref{Eq04}. At the same time, the electric field operator can be compactly expressed as:
\begin{equation}
\hat{\boldsymbol{E}}^{(+)}({\bf r};\omega)=i\omega\mu_0\int{d^3{\bf r}'{\bf G}({\bf r},{\bf r}',\omega)\cdot\hat{\bf j}({\bf r}';\omega)},
\label{Eq06}
\end{equation}
where $\hat{\bf j}({\bf r};\omega)$ stands for the current density operator. This latter expression, relating the EM fields with their sources (i.e., the electric current density operator), justify the relationship of the emission spectra and the quantum version of the~FDT. Therefore, inasmuch as the dynamics of the model is ruled by the Hamiltonian, the explicit expressions for the EM current density operator are ultimately determined by the dynamic of the polaritonic operators. Specifically, working under the Heisenberg picture, the dynamical behavior of the polaritonic operators characterizing the system is dictated by the Heisenberg equation of motion:
\begin{equation}
i\hbar\partial_t\hat{\boldsymbol{\mathcal{O}}}\equiv\left[\hat{\boldsymbol{\mathcal{O}}},\hat{\mathcal{H}}\right].
\label{Eq07}
\end{equation}
Regarding the polaritonic operators, they should obey the canonical equal-time commutation relations: $[\hat{\bf f}({\bf r},\omega_f;t),\hat{\bf f}({\bf r}',\omega_f';t)]=[\hat{\bf f}^\dagger({\bf r},\omega_f;t),\hat{\bf f}^\dagger({\bf r}',\omega_f';t)]=0$, and $[\hat{\bf f}({\bf r},\omega_f;t),\hat{\bf f}^\dagger({\bf r}',\omega_f';t)]=\hat{\mathbb{I}}\delta{[{\bf r}-{\bf r}']}\delta{[\omega_f-\omega_f']}$, where $\hat{\mathbb{I}}$ is the identity operator. So, for the free-evolving term it follows that:
\begin{equation}
i\hbar\partial_t\hat{\bf f}_{\rm 0}=[\hat{\bf f},\hat{\mathcal{H}}_{\rm 0}]=\hbar\omega_f\hat{\bf f}({\bf r},\omega_f;t).
\label{Eq08}
\end{equation}
Likewise, the time-modulated contribution is given by:
\begin{align}
\!\!\!\!\nonumber i\hbar\partial_t\hat{\bf f}_{\rm T}&=[\hat{\bf f}({\bf r},\omega_f;t),\hat{\mathcal{H}}_{\rm T}]=\int{d^3{\bf r}'\left[\hat{\boldsymbol{\mathcal{P}}}({\bf r}';t)\hat{\boldsymbol{\mathcal{E}}}({\bf r}';t),\hat{\bf f}({\bf r},\omega_f;t)\right]}\\
&=\int{d^3{\bf r}'\int_{0}^{t}{d\tau \Delta\chi({\bf r}',t,\tau)\left[\hat{\boldsymbol{\mathcal{E}}}({\bf r}';\tau)\hat{\boldsymbol{\mathcal{E}}}({\bf r}';t),\hat{\bf f}({\bf r},\omega_f;t)\right]}}.\!\!\!\!
\label{Eq09}
\end{align}
And similarly for the Hermitian conjugated operators $\hat{\bf f}_{\rm 0}^\dagger$ and~$\hat{\bf f}_{\rm T}^\dagger$. Inserting these expressions respectively into \eqref{Eq04}, and subsequently using \eqref{Eq06}, leads to the free-evolving and time-modulated contributions of both the fluctuating electric field and current density operators. This general procedure essentially conforms the basis to calculate the correlations and hence the thermal emission spectra in time-varying media~\cite{VazquezLozan02023A}.

This formalism has thus far only been assessed under the sharply simplistic, though widespread, assumption whereby the susceptibility exhibits a non-dispersive (i.e., temporally-local or instantaneous) time-modulation profile:
\begin{equation}
\Delta\chi({\bf r},t,\tau)=\Delta\tilde{\chi}({\bf r},t)\delta{[t-\tau]}.
\label{Eq10}
\end{equation}
Despite being an unphysical assumption, as it assumes the existence of a memory-less system responding instantaneously to the time-modulation, it has been proved to be useful for circumventing the rather intricate form of time-dependent quantum commutation relations~\cite{Vogel}, providing with reasonable~predictions in the limit of sufficiently low frequencies~\cite{VazquezLozan02023A,VazquezLozan02023B}. However, as schematically depicted in~\hyperref[Fig_01]{Fig.~1(b)}, within the context of thermal light emission, it has the pernicious effect of leading to a spectral divergence in the high-frequency limit [see solid green curve in \hyperref[Fig_01]{Fig.~1(b)}]. This realization can be readily understood from two reasons: (1) the Fourier transform of the Dirac delta function in time-domain is a constant real-valued function for all frequencies [see dashed red curve in \hyperref[Fig_01]{Fig.~1(b)}], and (2) the vacuum energy spectrum, $\hbar\omega/2$, turns out to be comparable and even higher than the black-body thermal spectrum, $\hbar\omega\Theta(\omega,T)$ [see solid black curve in \hyperref[Fig_01]{Fig.~1(b)}], in a relatively high-frequency limit, even at MIR frequencies tied to room temperatures~\cite{Liberal2024}. Then, given that the time-modulation is a mechanism to dynamically amplify these zero-point quantum vacuum fluctuations, it brings about a monotonously increasing contribution at higher frequencies. Therefore, in absence of a dispersive response in the time-modulation that may yield a frequency-tempered function, the emission spectrum shall display a diverging behavior in the high-frequency limit.

\begin{figure}[t!]
	\centering
	\includegraphics[width=0.985\linewidth]{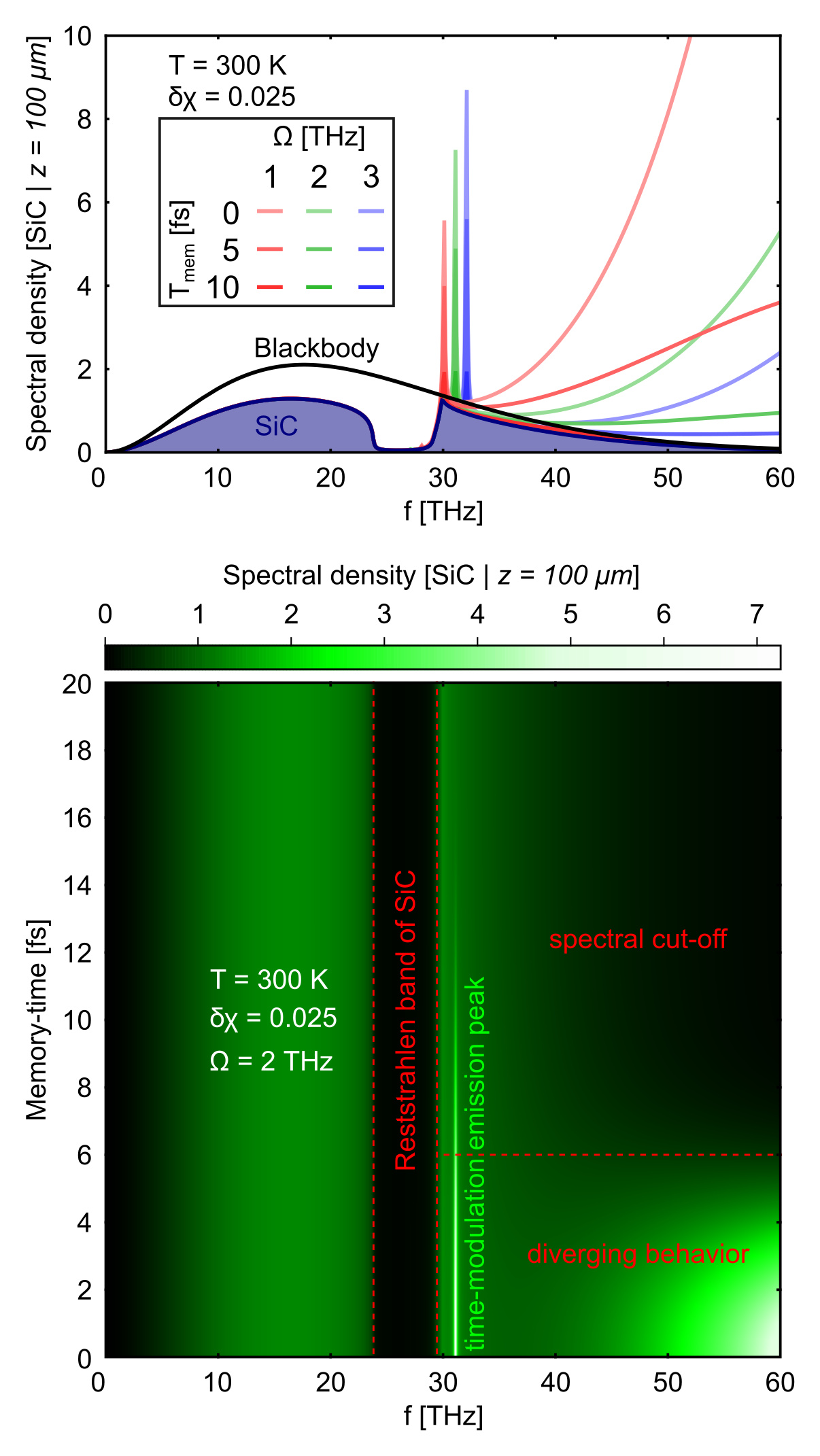}
	\caption{Dispersion effects in thermal emission of time-modulated media. (a)~Emission spectra of a semi-infinite planar slab made of SiC at room temperature ($T=300$ K) with a time-varying susceptibility subjected to a time-harmonic modulation with $\delta\chi=0.025$ for different values of modulation frequency ($\Omega$) and memory-time ($T_{\rm mem}$). As shown for each value of $\Omega$, the larger the memory-time the sharper the cut-off at the high-frequency limit, thereby explicitly showing the effects of dispersion on the emission spectra. For clarity, stationary (non-time-modulated) SiC (blue area) and blackbody spectrum (black curve) are shown. (b)~Color map representing different regimes in terms of the memory-time, which ultimately characterize the dispersive time-modulation of the material.}
	\label{Fig_02}
\end{figure}

In order to extend the theoretical framework beyond this ideal approach toward a more realistic scenario including dispersive time-modulations, we reconsider the above expression for the time-dependent electric field tied to the polarization field operator [see~\eqref{Eq04}], recasting it in terms of a Taylor series expansion:
\begin{equation}
\!\!\!\!\hat{\boldsymbol{\mathcal{E}}}^{(+)}({\bf r};\tau)\!=\!\int{d^3{\bf r}'\!\!\int{d\omega_f {\bf G}_{\rm E}({\bf r},{\bf r}',\omega_f)\cdot\hat{\bf f}({\bf r}',\omega_f;t) e^{-i\omega_f(\tau-t)}}}.\!\!\!\!
\label{Eq11}
\end{equation}
Upon this ground, instead of using the Dirac delta function, we can now model the time-varying susceptibility by considering its analytical extension regarded as a Gaussian distribution, that~is:
\begin{equation}
\Delta\chi({\bf r}',t,\tau)=\Delta\tilde{\chi}({\bf r}',t)\frac{H{[t-\tau]}}{T_{\rm mem}\sqrt{\pi}}e^{-(t-\tau)^2/T_{\rm mem}^2},
\label{Eq12}
\end{equation}
where $H{[t-\tau]}$ is the Heaviside step function, and $T_{\rm mem}$ is the temporal width defining the memory of the time-modulation [see the schematic representation in~\hyperref[Fig_01]{Fig.~1(c)}]. From these expressions it can be demonstrated that the Heisenberg equation of motion for the time-modulated contribution of the polaritonic operators [see~\eqref{Eq09}] can be compactly expressed as:
\begin{align}
&\!\!\nonumber i\hbar\partial_t\hat{\bf f}_{\rm T}({\bf r},\omega_f;t)=[\hat{\bf f}({\bf r},\omega_f;t),\hat{\mathcal{H}}_{\rm T}]\\
&\!\!\approx-e^{-\omega_f^2T_{\rm mem}^2/4}\!\int\!{d^3{\bf r}'\Delta\tilde{\chi}({\bf r}',t){\bf G}_{\rm E}^*({\bf r}',{\bf r},\omega_f)\hat{\boldsymbol{\mathcal{E}}}({\bf r}';t)\Xi_{T_{\rm mem}}(t)},\!\!
\label{Eq13}
\end{align}
where $\Xi_{T_{\rm mem}}(t)={\rm Erf}{[t/T_{\rm mem}]}$, i.e., the so-called Gauss error function. It is worth remarking that this relatively simple expression has been obtained under the condition that $T_{\rm mem}\omega_f\to 0$, which still restricts the scope of the formalism to systems with very short, though finite, memory times. At any rate, it generalizes previous results and allows to recovering the non-dispersive (instantaneous) approximation leading to the case of a memory-less system. Indeed, it is straightforward to prove that, by construction, in the limit of zero memory, i.e., $T_{\rm mem}\to 0$, the time-varying susceptibility becomes, $\Delta\chi({\bf r}',t,\tau)\to\Delta\tilde{\chi}({\bf r}',t)\delta{[t-\tau]}/2$ (notice that the Dirac delta function, regarded as a broad Gaussian kernel centered at the time $\tau$, only account for one half of the whole area, leaving aside the other part of the function). Likewise, concerning the dynamic of the time-modulated polaritonic operators~[cf.~\eqref{Eq13}], the memory-less limit leads to $\Xi_{T_{\rm mem}\to0}(t)\to 1$, so that $i\hbar\partial_t\hat{\bf f}_{\rm T}({\bf r},\omega_f;t)\to-\int{d^3\tilde{\bf r}}\Delta\tilde{\chi}(\tilde{\bf r},t)\bar{\bf G}_{\rm E}(\tilde{\bf r},{\bf r},\omega_f)\hat{\boldsymbol{\mathcal{E}}}(\tilde{\bf r};t)$, thus bringing forth the non-dispersive dynamical behavior~\cite{VazquezLozan02023A}.

Noteworthily, this approach can be seamlessly incorporated into the non-dispersive formalism as a straightforward mathematical extension. Indeed, since the exponential factor preceding the integral in \eqref{Eq13} is independent of time, it can be taken out of the time integral that yields the expression of time-modulated polaritonic operators. As a result, for both the field and current density operators, this contribution appears merely as a multiplicative factor that accompanies all previously derived expressions for temporal modulation. Therefore, the time-modulated contribution to the emission spectra remains the same as in the non-dispersive case, but now being simply weighted by the square of the aforementioned factor, i.e., $e^{-\omega_f^2T_{\rm mem}^2/2}$, accounting for the correlation of two time-modulated electric field operators.

For the sake of comparison, we explicitly illustrate the practical consequences of introducing the dispersion into the theoretical model by considering the same explanatory example presented in Ref.~\cite{VazquezLozan02023A}. It consists of a semi-infinite planar slab made of silicon carbide (SiC) ($z<0$), at room temperature~($300$~K), in contact with vacuum ($z>0$). Here, the SiC substrate is optically characterized by the Drude-Lorentz permittivity, $\varepsilon(f)=\varepsilon_{\infty}(f_{\rm L}^2-f^2-i\gamma f)/(f_{\rm T}^2-f^2-i\gamma f)$, with $\varepsilon_\infty=6.7$, $f_L=29.1$~THz, $f_T=23.8$~THz, and $\gamma=0.14$~THz~\cite{Shchegrov2000}, being, respectively, the high-frequency-limit permittivity, the longitudinal and transverse optical phonon frequencies, and the damping factor. In addition, to characterize the modulation we assume that the time-varying susceptibility~of the SiC is subjected to a time-harmonic profile: $\Delta\tilde{\chi}({\bf r},t)=\varepsilon_0\delta\chi\sin{\Omega t}$ [see~\hyperref[Fig_01]{Fig.~1(a)}]. Upon this ground,~in~\hyperref[Fig_02]{Fig.~2}, we present the analytical results of the far-field [$z=100$ $\mu$m] emission spectra of this time-modulated dispersive system for different values of the frequency of modulation, $\Omega$, and memory-time,~$T_{\rm mem}$. Specifically, in~\hyperref[Fig_02]{Fig.~2(a)}, we show the effect of the dispersion, characterized in terms of the memory-time ($T_{\rm mem}$), for different values of modulation frequency ($\Omega$). As can be observed, regardless of the value of modulation frequency $\Omega$, higher dispersion (i.e., longer $T_{\rm mem}$) leads to sharper cut-off frequencies. Likewise, the intensity of the ENZ-induced emission peak reduces as dispersion effects get dominant. Particularizing into the case of $\Omega=2$ THz, in~\hyperref[Fig_02]{Fig.~2(b)} we represent the emission spectra for a continuum of $T_{\rm mem}$. This highlights key features of the system and reveals two distinct characteristic regimes -- one exhibiting a diverging spectral behavior and other yielding the occurrence of spectral cut-offs -- both determined by the underlying dispersion.

In conclusion, we have theoretically explored the effects of introducing temporal dispersion into the quantum formalism of thermal emission in time-varying media. This approach essentially raises on the introduction of a kernel response function associated with time modulation that accounts for a memory-time, meaning that the material response is not instantaneous. Our analysis reveals that the main consequence of this realization is the occurrence of a high-frequency cut-off, which not only eliminates the non-physical divergence but also enables more accurate outcomes. Therefore, this work lays the foundation for calculating the total power radiated by time-modulated materials and opens up opportunities to investigate the intrinsic properties of materials in response to external temporal modulations. In this vein, beyond its fundamental significance, we hope that our findings might provide valuable insights for designing experimental setups to observe the effects of thermal emission in time-varying media.

\newpage

\noindent{\textbf{\Large Acknowledgment.}} This work was supported by ERC Starting Grant No. ERC-2020-STG-948504-NZINATECH. A.V.-C. acknowledges support from Beca de Colaboraci\'on provided by Universidad P\'ublica de Navarra (Resoluci\'on No. 2316/2023). I.L. acknowledges support from Ram\'on y Cajal fellowship RYC2018-024123-I. J.E.V.-L. further acknowledges support from Juan de la Cierva--Formaci\'on fellowship FJC2021-047776-I.\\

\noindent{\textbf{\Large Disclosures.}} The authors declare no conflicts of interest.

% Bibliography

%\bibliography{sample}

%\bibliographyfullrefs{sample}

%Manual citation list

\end{document}